\documentclass[prx,aps,twocolumn,superscriptaddress]{revtex4}

\usepackage{color}
\usepackage[pdftex]{graphicx}

\usepackage{dcolumn}
\usepackage{bm}
\usepackage{amssymb}

\begin{document}


\title{Theory of magnetotransport in artificial kagome spin ice}

\author{Gia-Wei Chern}
\affiliation{Department of Physics, University of Virginia, Charlottesville, VA 22904, USA}

\date{\today}

\begin{abstract}
Magnetic nanoarray with special geometries exhibits nontrivial collective behaviors similar to those observed in the spin ice materials. Here we present a novel circuit model to describe the complex magnetotransport phenomena in artificial kagome spin ice. In this picture, the system can be viewed as a resistor network driven by voltage sources that are located at vertices of the honeycomb array. The differential voltages across different terminals of these sources are related to the ice-rules that govern the local magnetization ordering. The circuit model relates the transverse Hall voltage of kagome ice to the underlying spin correlations. Treating the magnetic nanoarray as metamaterials, we present a mesoscopic constitutive equation relating the Hall resistance to magnetization components of the system. We further show that the Hall signal is significantly enhanced when the kagome ice undergoes a magnetic-charge ordering transition. Our analysis can be readily generalized to other lattice geometry, providing a quantitative method for the design of magnetoresistance devices based on artificial spin ices. 
\end{abstract}

\maketitle

\section{Introduction}

Artificial spin ice~\cite{nisoli13,gilbert16} as a tailorable metamaterial has recently attracted considerable attention due to its potential technological applications. These metamaterials are nano-structured planar arrays of single-domain ferromagnetic wires that behave like mesocsopic Ising spins~\cite{wang06,tanaka06,qi08}. The special geometrical arrangement of nanowires in the array leads to frustrated interactions and disordered Ising spins even at temperatures well below the energy scale of nearest-neighbor couplings. However, unlike the simple disordered paramagnet, spin ice at low temperatures is a highly correlated state in which local arrangement of spins is governed by the so-called ice rules, similar to the local proton ordering in water ice~\cite{bernal33}. 
For example, the ice-rules in the square ice array dictates that the low-energy magnetic states of a vertex are the six two-in-two-out configurations~\cite{wang06}. 
Moreover, high-energy vertices that violate the ice rules are topological defects carrying a net magnetic charge, hence behaving like emergent magnetic monopoles~\cite{castelnovo08}.

\begin{figure}[t]
\includegraphics[width=0.91\columnwidth]{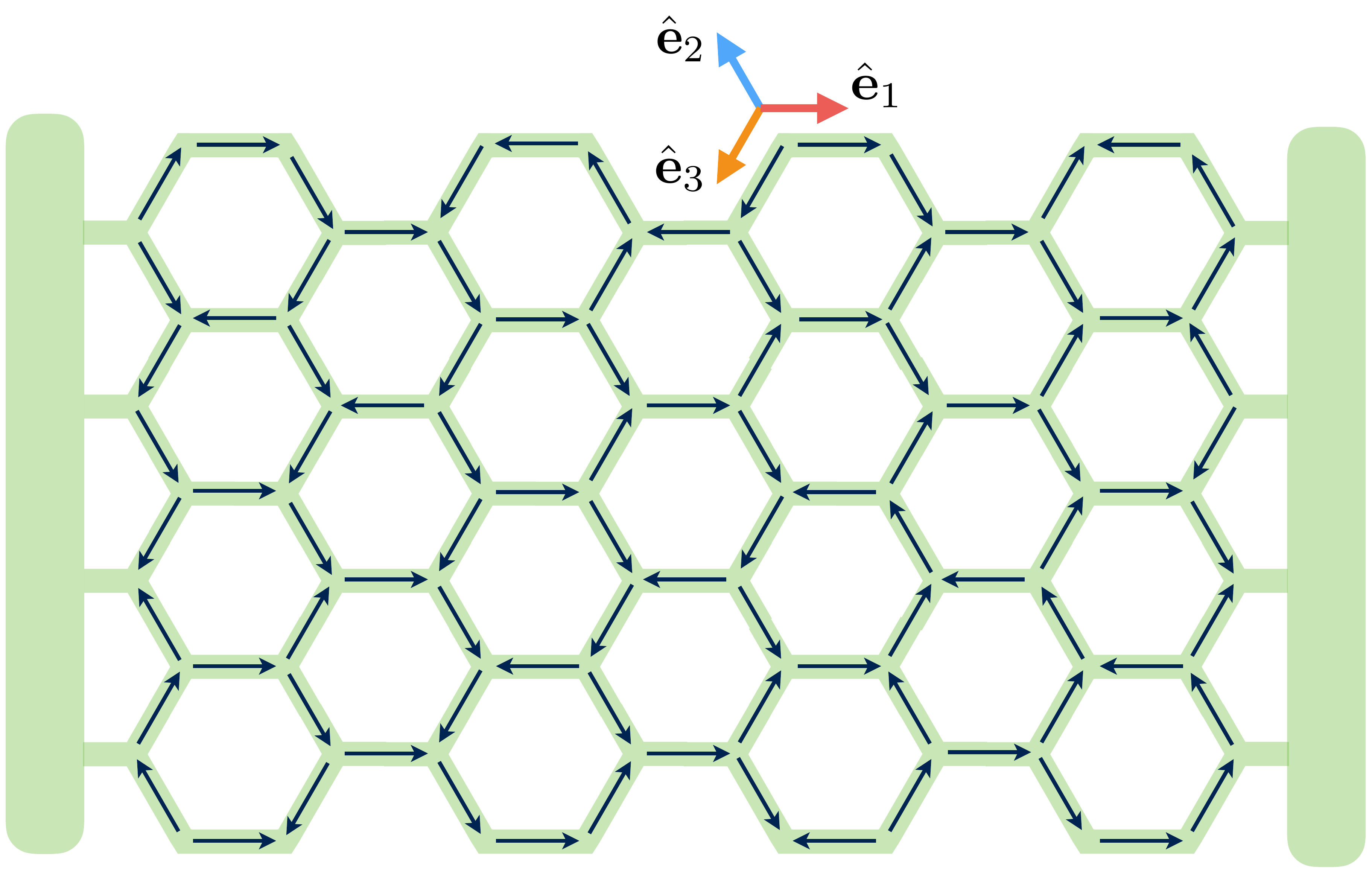}
\caption{(Color online) Artificial kagome spin ice consisting of array of nanoscale magnetic wires. The unit vectors $\hat{\mathbf e}_{1, 2, 3}$ are along the long axes of wires in the three sublattices. Magnetization of a wire is specified by a Ising variable. The Ising variables, which sit at centers of wires, form a kagome lattice.
\label{fig:honeycomb}}
\end{figure}

Magnetic charge degrees of freedom play a prominent role in the physics of spin ice materials. In particular,  artificial spin ices containing vertices of odd coordination numbers exhibit novel phenomena that can be attributed to the existence of the uncompensated background magnetic charges~\cite{moller09,chern11,schumann10,rougemaille11,gilbert14,chern16,farhan16}. A remarkable example is the kagome spin ice in which a spontaneous ordering of magnetic charges precedes the long-range spin order~\cite{moller09,chern11}. This exotic magnetic phase has recently been observed in both artificial and natural kagome ice materials~\cite{zhang13,dun16,paddison16}. Over the years, artificial kagome spin ice, in which the nanowires are arranged in a hexagonal network shown in Fig.~\ref{fig:honeycomb}, has served as an experimental platform for investigating intriguing phenomena related to magnetic charges and geometrical frustration~\cite{hugli12}. 

In addition to magnetic behaviors, the magnetotransport properties of metallic kagome ice have been studied in recent experiments~\cite{tanaka06,branford12,le15,le16,park17}. The observed  Hall voltage can be attributed to the anisotropic magnetoresistance (AMR) effect. Interestingly, unlike the AMR behaviors seen in bulk ferromagnets, the field induced abrupt changes in Hall signals of artificial spin ice are related to collective many body phenomena~\cite{le16}. 
A rather intriguing result is the observation of a large Hall voltage even at very small magnetic field. Crucially, as AMR effect of individual wires vanishes in this zero field limit, the observed finite Hall voltage was shown to result from the vertex regions of the honeycomb network~\cite{le16}. 

In~this paper, we develop an effective circuit model to understand the complex magnetotransport phenomena in artificial kagome spin ice. While nanowires are treated as simple resistors in our model, individual vertex acts as a voltage source whose parameters are controlled by local spin correlations. The circuit model not only underscores the many-body origin of the Hall signal in kagome ice, but also provides a guiding principle for designing magnetoresistive devices based on these metamaterials.  By combining extensive Monte Carlo simulations with the circuit model, we present a simple empirical formula that relates the transverse Hall voltage of kagome ice to its magnetization components. We further demonstrate that the Hall signal is enhanced by the ordering of magnetic charges in kagome spin ice.

\section{Circuit model}

We start with the AMR effect~\cite{mcguire75} that describes the dependence of resistivity on the angle enclosed by the magnetization and the current density vector:
\begin{eqnarray}
	\label{eq:AMR}
	\mathbf E = \rho_0 \mathbf J + \Delta \rho \left(\hat{\mathbf m} \cdot \mathbf J\right)\, \hat{\mathbf m}.
\end{eqnarray}
Here $\mathbf E$ is the electric field, $\mathbf J$ is the current density vector, $\hat{\mathbf m}$ is a unit vector pointing along the magnetization direction, $\rho_0$ is the isotropic bulk resistivity, and $\Delta \rho$ is the anisotropic magneto-resistivity. A voltage drop $V_{\perp} = \Delta \rho\, I \sin(2 \phi) / t$ transverse to the current flow develops whenever the current is neither perpendicular nor parallel to magnetization; here $\phi$ is the angle between the two vectors $\mathbf J$ and $\hat{\mathbf m}$ and $t$ is thickness of the film. This so-called planar Hall effect has been exploited for the design of magnetic field sensors~\cite{henriksen10}.

\begin{figure}[t]
\includegraphics[width=0.99\columnwidth]{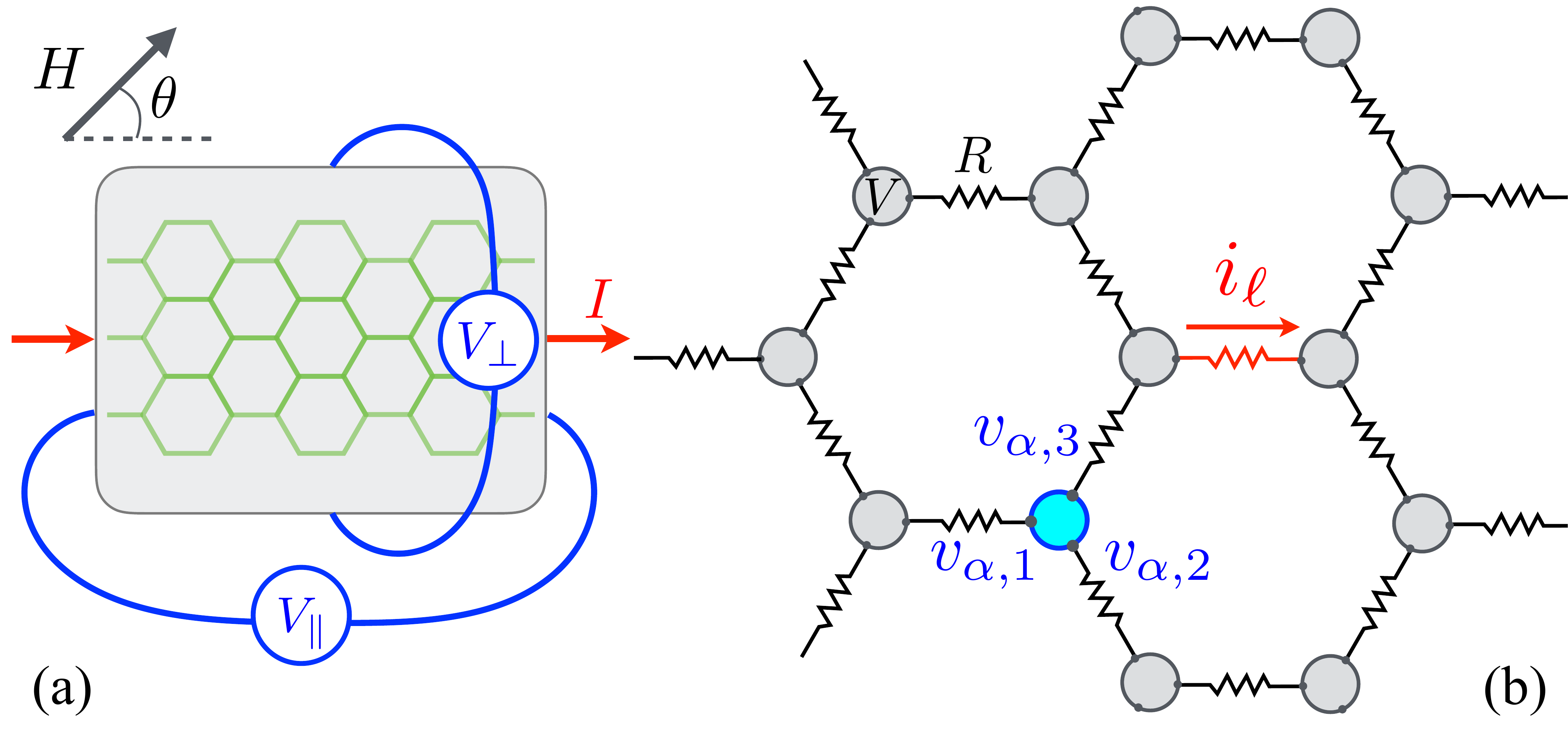}
\caption{(a) Experimental setup for measuring magnetotransport of the honeycomb array. (b) Effective circuit model for kagome spin ice. Each wire has a resistance $R$, and each vertex behaves as a three-terminal voltage source. The differential voltages are controlled by the local Ising spin ordering.
\label{fig:circuit}}
\end{figure}

We consider an experimental setup shown in Fig.~\ref{fig:circuit}(a) where current flow $I$ is parallel to the horizontal axis $\hat{\mathbf e}_1$ and an in-plane magnetic field $\mathbf H = H (\cos\theta, \sin\theta)$ is applied to the nano-array. In the absence of field, the magnetization of individual wires point along the corresponding long axes $\mathbf m_\ell = \sigma_\ell \, m_0 \,\hat{\mathbf e}_{s_\ell}$ due to a strong magnetostatic shape anisotropy; here $\sigma_\ell = \pm 1$ is an Ising variable, $s_\ell = 1, 2, 3$ denotes the sublattice of the wire, and $m_0$ is the saturation magnetization. It was shown that field-induced twisting of local magnetization away from the long axes is important for explaining the large-field magnetoresistance data~\cite{le16}. However, this is mainly the bulk AMR effect, and since we are interested in the small field regime, we will neglect such deformations and treat each nanowire as a Ising spin $\sigma_\ell$. On the other hand, even in the $H = 0$ limit the vertex region exhibits a complex magnetization profile depending on the Ising states of the three wires that are connected to the vertex under consideration. 

A complete calculation of the electric field and current density for a given magnetization profile of the vertex is a difficult task, which often requires numerical computations. One has to solve the electrostatic equation $\nabla \cdot \mathbf E = 0$ and the continuity equation $\nabla \cdot \mathbf J = 0$ subject to the appropriate boundary conditions along with the AMR constitution equation~(\ref{eq:AMR}). Here we assume that $\Delta \rho$ is small compared with the isotropic resistivity $\rho_0$ and employ a perturbation approach. Basically, we expand both field and current as power series of the small parameter:  $\mathbf E = \mathbf E^{(0)} + \mathbf E^{(1)} + \cdots$ and $\mathbf J = \mathbf J^{(0)} + \mathbf J^{(1)} + \cdots$, where $\mathbf E^{(n)}$ and $\mathbf J^{(n)}$ are proportional to $\Delta\rho^n$. A perturbative solution is obtained by substituting these expansions into Eq.~(\ref{eq:AMR}) and solve the resultant equation order by order. The leading zeroth order result is the simple Ohm's law $\mathbf E^{(0)} = \rho_0 \,\mathbf J^{(0)}$. In other word, the kagome spin ice at this leading order is simply a honeycomb resistor network with each nanowire having a resistance $R = \rho_0 l / w t$, where $l$, $w$, and $t$ are the length, width, and thickness, respectively, of the wire. Apparently, there is no transverse voltage $V^{(0)}_{\perp} = 0$ at this leading order.

\begin{figure}[t]
\includegraphics[width=0.99\columnwidth]{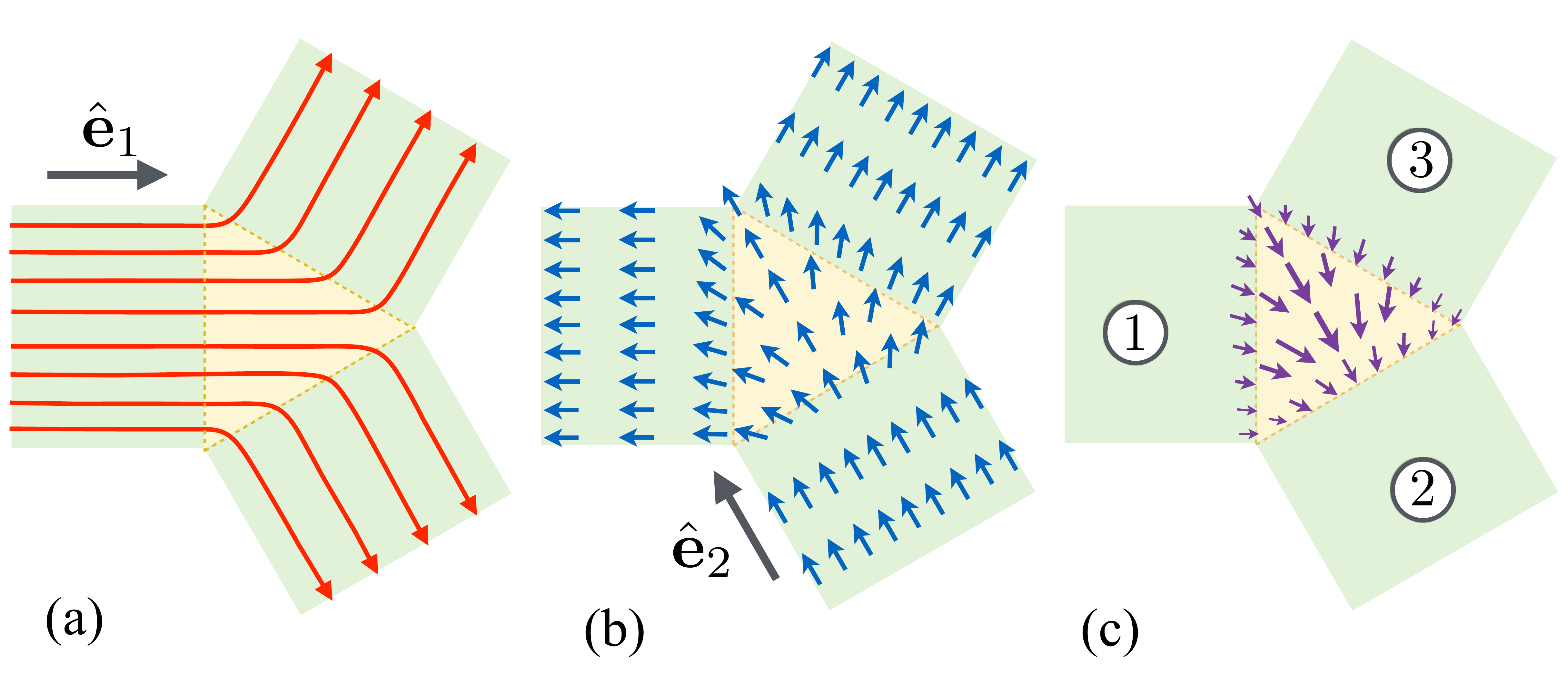}
\caption{(a) Current density $\mathbf J$, (b) magnetization profile $\mathbf m(x, y)$, and (c) AMR induced first-order electric field $\mathbf E^{(1)}$ around a vertex region with a one-in-two-out Ising configuration: $\sigma_2 = +1$, $\sigma_1 = \sigma_3 = -1$. A nonzero magnetic charge density $\rho_m = \nabla \cdot \mathbf m$ accounts for the net magnetic charge of this vertex. 
\label{fig:vertex}}
\end{figure}

\subsection{voltage sources}
\label{sec:v_source}

\begin{figure}[t]
\includegraphics[width=0.85\columnwidth]{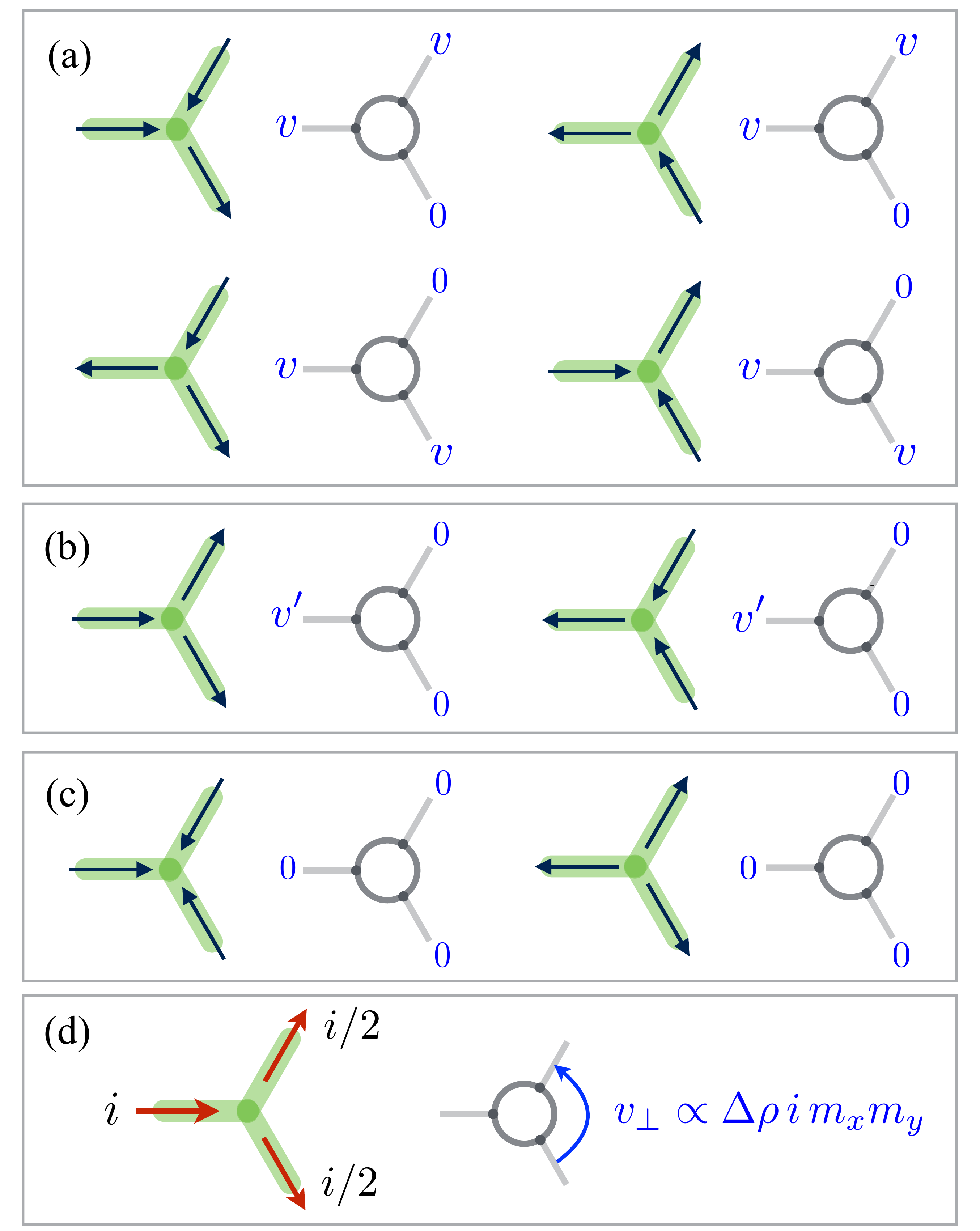}
\caption{The eight different vertex configurations in kagome spin ice and their corresponding voltage source representation in circuit model. The vertices are classified into three categories: (a) vertices contributing to both longitudinal and transverse voltages, (b) vertices showing only longitudinal voltage drop, and (c) no voltage difference between the three terminals. Vertices in categories (a) and (b) satisfy the 2-in-1-out or 1-in-2-out ice rules.  Here we only show types of vertices belonging to one sublattice; the other sublattice can be straightforwardly obtained using similar analysis. Panel (d) emphasizes that the voltage source representation is obtained based on a current flow from left to the right in the network. Also shown is the contribution $v_{\perp}$ of the vertex to the transverse Hall voltage of the system.
\label{fig:v_source}}
\end{figure}

Next we turn to the first-order solution. The electric field is readily obtained: $\mathbf E^{(1)} = \Delta \rho \, (\hat{\mathbf m} \cdot \mathbf J^{(0)} )\, \hat{\mathbf m}$, which depends on the zeroth-order current density. Since in the low field limit, magnetization is aligned with the long axis of the wires, which is also the direction of $\mathbf J^{(0)}$, the first-order solution only introduces a correction to the wire resistance $R = (\rho_0 + \Delta \rho) l / wt$. On the other hand, the magnetization has a complex profile $\mathbf m(x,y)$ in the vertex region and in general does not point along the current direction; as demonstrated in the example shown in Fig.~\ref{fig:vertex}. Here the vertex is in a one-in-two-out state, with the minority spin (the `in' spin) being in wire $\textcircled{2}$.  In this case, while the current density $\mathbf J^{(0)}$ is parallel to the $\hat{\mathbf e}_1$ direction in the vertex triangular area, the magnetization vector $\mathbf m$ mainly points toward the $\hat{\mathbf e}_2$ direction in the vertex region. The resultant first-order electric field is then $\mathbf E^{(1)} \approx -J \Delta\rho\, \hat{\mathbf e}_2$; see Fig.~\ref{fig:vertex}(c). As a result a voltage drop $v \approx \alpha \Delta\rho \, i /t$ develops along the negative $\hat{\mathbf e}_2$ axis across the vertex area, where $i$ is the current in a wire and $\alpha$ is a numerical constant. At first order, this 1-in-2-out vertex can be viewed as a three terminal voltage source imposing a voltage drop $v$ between terminals of majority wires (the `out' spins) and the minority one; see Fig.~\ref{fig:v_source}(a).

Similar conclusion can be reached for vertices of different configurations. Three distinct voltage sources and their corresponding vertex configurations are summarized in Fig.~\ref{fig:v_source}. We note that the time-reversed version of these vertices, i.e. $\sigma_\ell \to -\sigma_\ell$, gives rise to exactly the same voltage source since the AMR relation Eq.~(\ref{eq:AMR}) is invariant under the transformation $\mathbf m \to -\mathbf m$. In Fig.~\ref{fig:v_source}(b) where the horizontal wire is the minority spin, the vertex magnetization is parallel or antiparallel to the current density vector, giving rise to a voltage drop $v' \approx 2 v$. On the other hand, the 3-in or 3-out vertices that violate the ice rule [case (c)] do not introduce any voltage differences among the three terminals. Consequently, the presence of such higher-energy vertices or magnetic monopoles reduces the Hall signal. Finally, it is worth noting that since the overall current $I$ is along the horizontal direction, the nontrivial transverse Hall signal in kagome spin ice mainly comes from ice-rule vertices shown in Figs.~\ref{fig:v_source}(a).

\begin{figure}[t]
\includegraphics[width=0.95\columnwidth]{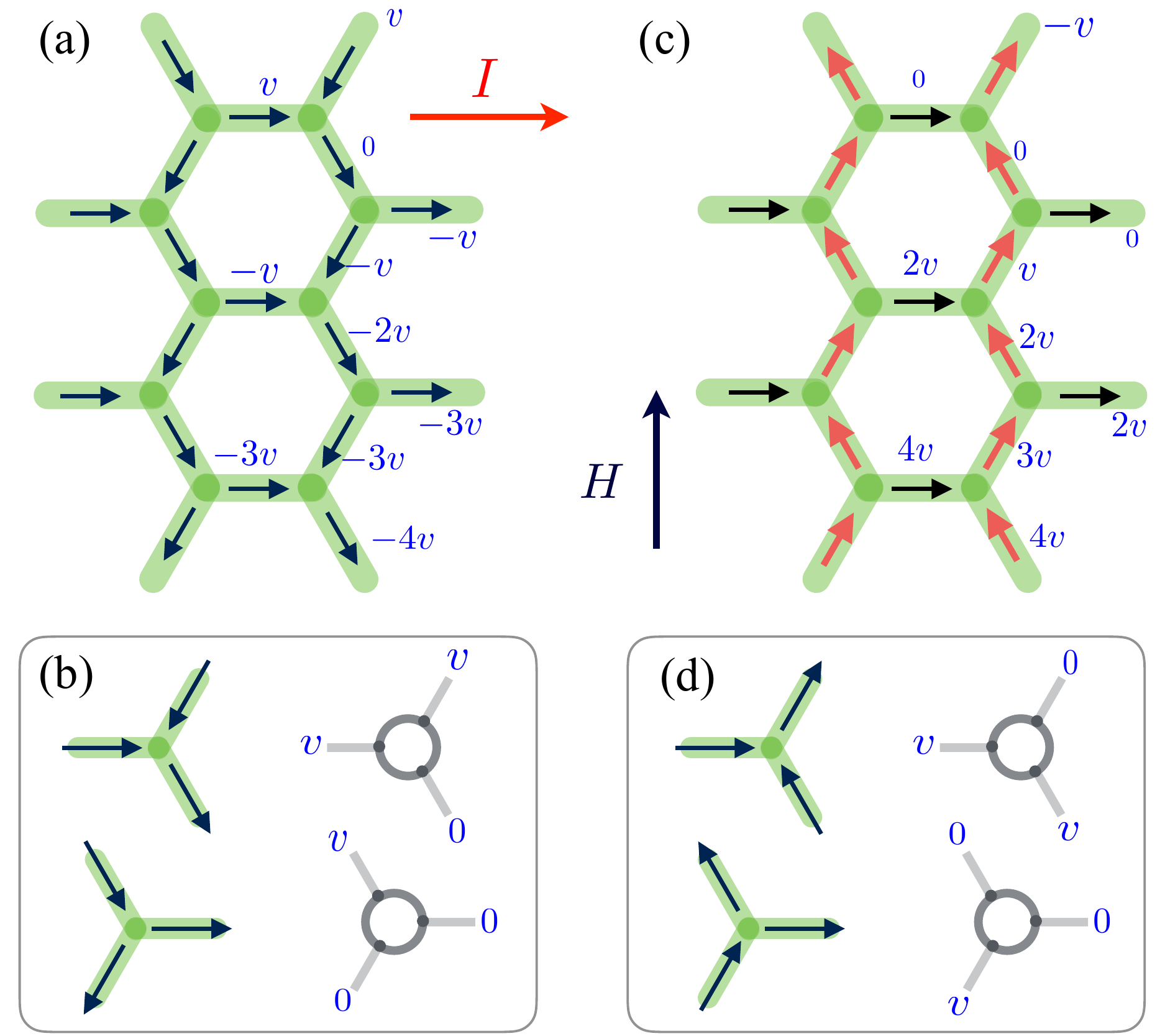}
\caption{(a) A long-range ordered configuration of the kagome ice with all non-horizontal spins pointing downward and horizontal ones pointing to the right. This state exhibits a maximum Hall voltage $V_{\perp} = 2 L_y v$, where $L_y$ is the linear size of the array in the $y$ direction. The constituent vertices and their corresponding voltage source representations are shown in (b). (c) An ordered state obtained from (a) by flipping all non-horizontal spins. The corresponding Hall voltage also changes sign: $V_{\perp} = - 2 L_y v$. (d) the vertices and voltage sources appearing in state (c).
\label{fig:m-rev-y}}
\end{figure}

As a simple application, we consider a kagome ice with long range spin order shown in Fig.~\ref{fig:m-rev-y}(a). In this configuration, all horizontal spins point to the right while non-horizontal ones point downward. This ordered ice state is comprised of vertices shown in Fig.~\ref{fig:m-rev-y}(b). Using the table of voltage source representation shown in Fig.~\ref{fig:v_source}, one can assign a voltage to each link and easily show that the system develops a Hall voltage $V_{\perp} = 2 L_y v$ and a longitudinal voltage $V_{\parallel} =- 2 L_x v$, where $L_x$, $L_y$ are the linear sizes of the array in the $x$, $y$ directions, respectively. Interestingly, both voltages are inverted when the non-horizontal spins are flipped by a large magnetic field along the $y$ direction; see Fig.~\ref{fig:m-rev-y}(b). As will be discussed in Sec.~\ref{sec:m-reversal}, the reversal of the non-horizontal spins is a collective behavior similar to the avalanche process. Such sudden change of Hall voltage during magnetization reversal has been demonstrated experimentally~\cite{le16}. Our circuit model provides a simple explanation for this phenomenon.

The method of voltage source representation is a simplified version of the line-integral method discussed in Ref.~\cite{le16} for estimating the Hall voltage of ASI. The line-integral method is very useful when combined with the micromagnetic simulations. For a given magnetization profile $\mathbf m(x, y)$ obtained from the micromagnetic simulations, the first-order electric field induced by AMR is $\mathbf E^{(1)} = \Delta \rho (\mathbf m \cdot \mathbf J) \,\mathbf m$ as discuss above. Using a appropriate current density $\mathbf J(x, y)$ for the network, one can then estimate voltage drop across a given path $C$ as $\Delta V = \int_C \mathbf E^{(1)} \cdot d {\mathbf l}$. In particular, the transverse voltage $V_\perp$ of the system is obtained by considering a path spanning from the top to the bottom of the array. However, this is not a self-consistent perturbation calculation since it neglects the first-order correction in current, as discussed in the next section.



\subsection{loop currents and self-consistent circuit equations}

\begin{figure}[t]
\includegraphics[width=0.95\columnwidth]{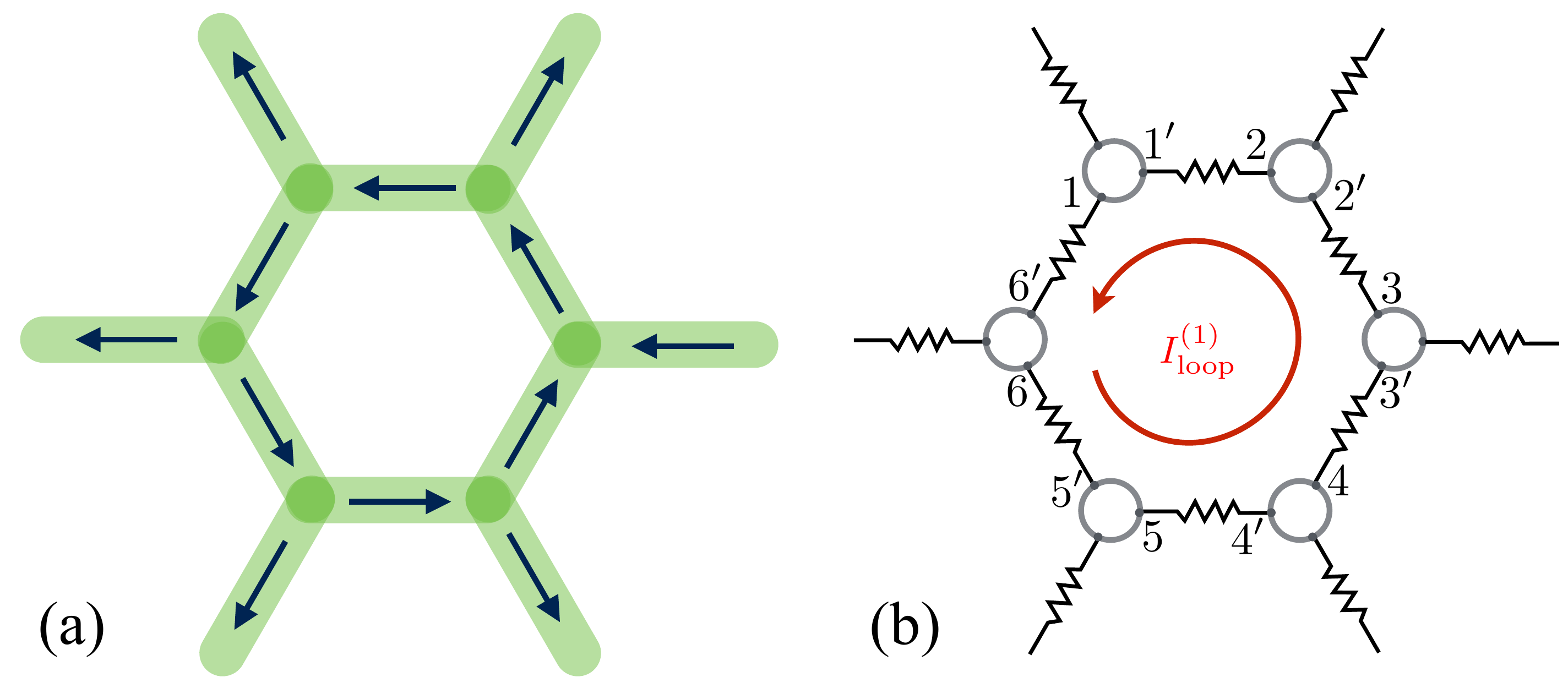}
\caption{(a) Configuration of spins around a hexagon loop and (b) the corresponding voltage source representation. Using rules of the voltage source summarized in Fig.~\ref{fig:v_source}, one can easily compute the voltage drop around the loop $\Delta V = \sum_{n=1}^6 V_{nn'}  = v' + v + v + (-v') + (-v) + v = 2v$. This indicates a finite first-order loop current $I^{(1)}_{\rm loop} = 2v / R$.
\label{fig:loop-curr}}
\end{figure}

In the examples shown in Fig.~\ref{fig:m-rev-y}, the AMR-induced voltages are assigned to individual links based on the assumption that there is no additional current flow in the network. This is possible for simple perfectly ordered states such as those shown in Fig.~\ref{fig:m-rev-y} in which the AMR-induced voltage changes add up to zero when transversing an arbitrary loop. As demonstrated in Fig.~\ref{fig:loop-curr}, for arbitrary spin configurations, the voltage drop around a loop is usually finite, implying a nonzero first-order correction in the form of a loop current. In general, a loop current exists whenever the line integral $\oint_L \mathbf E^{(1)} \cdot d{\bf l} \neq 0$. This also means that the simple voltage-source or the line-integral method of Ref.~\cite{le16} cannot produce consistent result in the presence of loop currents.

To obtain a self-consistent first-order solution induced by AMR, one thus needs to take into account the correction to the current density. However, a microscopic calculation based on the constitutive equation $\mathbf J^{(1)} = \mathbf E^{(1)} / \rho_0$ is rather cumbersome and not feasible for large scale simulations. A simple consistent approach is to solve the circuit equations for the first-order voltage and current corrections simultaneously. Here we discuss the setup of a complete linear equation system describing a resistor network driven by the voltage sources described in Fig.~\ref{fig:v_source}. We use $v_{\alpha}$ to denote the first-order voltage of vertex $\alpha$ and $i_\ell$ for the current of wire $\ell$.  These variables are related through equations:
\begin{eqnarray}
	\label{eq:linear1}
	i_{\alpha, 1} + i_{\alpha, 2} + i_{\alpha, 3} = 0, \\
	\label{eq:linear2}
	v_{\alpha, s} - v_{\beta, s} = i_{(\alpha\beta)} R.
\end{eqnarray}
The first equation is the Kirchhoff's current rule, which is simply a consequence of charge conservation. Here $i_{\alpha, s}$ denotes the current flowing through terminal $s$ of vertex~$\alpha$. The second equation is the Ohm's law for a wire of sublattice $s$ that connects a pair of vertices $(\alpha\beta)$.  The voltages at the three terminals of a vertex $\alpha$ are given by 
\begin{eqnarray}
	v_{\alpha, s} = v_\alpha + \mathcal{V}_s(\sigma_1,\sigma_2,\sigma_3), 
\end{eqnarray}
where $s = 1, 2, 3$ and $\sigma_{1,2,3}$ denote the Ising spins of the three wires connected to $\alpha$. The source function $\mathcal{V}_s$ depends on the three Ising spins and is defined according to the rules shown in Fig.~\ref{fig:v_source}.  For a network consisting of $N$ wires and $M = 2N/3$ vertices, the total number of unknowns include $N$ current variables $i_{\ell}$ and $M$ voltages $v_\alpha$. These unknown variables can be consistently determined from the set of linear questions~(\ref{eq:linear1}) and~(\ref{eq:linear2}) which contains $M$ Kirchhoff current equations and $N$ Ohm's equations. Special care has to be taken for open boundary conditions of a finite array.

The circuit model approach is particularly powerful when combined with large-scale numerical methods, such as relaxation dynamics or Monte Carlo simulations. In such hybrid approach, snapshots of Ising configurations from the simulations are used to set up the circuit equations~(\ref{eq:linear1}) and~(\ref{eq:linear2}), which are then solved to give the instantaneous longitudinal and Hall voltages of the spin-ice system. In Monte Carlo simulations, often thousands of snapshots are needed in order to obtain a meaningful average. Consequently, an efficient method for computing the AMR-induced voltages is highly desirable. It is worth noting that although the computational cost of solving the linear system in general scales as $\mathcal{O}(N^3)$, where $N$ is the number of spins or links, the sparse-matrix algorithms provide a linear-scaling method that can be feasibly combined with the large-scale Ising-spin simulations. Specifically, here we employ an efficient LU decomposition method for sparse matrices~\cite{armadillo} to solve the linear system. This allows us to efficiently simulate arrays consisting of $N \sim 10^4$ spins.


\section{magnetization reversal}
\label{sec:m-reversal}

Next we combine the circuit model with the relaxation dynamics simulations~\cite{mellado10,chern14} to investigate the field dependence of Hall resistance during magnetization reversal of the kagome ice. In these simulations, the magnetic field is applied opposite to the majority spin direction in the initial state. As field is increased, inversion of Ising spin occurs only when the local field, which is the sum of external and dipolar fields projected onto the wire direction, exceeds a threshold:
\begin{eqnarray}
	-(\mathbf H_{\rm ext} + \mathbf H_{\rm dip}) \cdot \mathbf m_\ell > H_{c, \ell}.
\end{eqnarray}
Due to the quenched disorder of the nano-array, the wire coercive field $H_{c, \ell}$ is a random variable. For simplicity, we assume $H_{c, \ell}$ is sampled from a Gaussian distribution. As already discussed in previous studies~\cite{mellado10,chern14}, magnetization reversal in artificial spin ice proceeds in the form of avalanches. The magnitude of the avalanches is in general determined by the spreading of the random coercive fields $H_{c, \ell}$.
It is also worth noting that when the magnetic field is applied at an angle $\varphi$ to a wire, the Zeeman force along the long axis is reduced by a $\cos\varphi$ factor. Consequently, the effective coercive field of the wire is enhanced to $H_c^{\rm eff}(\varphi) = H_c / \cos \varphi$. 

For convenience, we present our numerical results using normalized quantities. The magnetizations are normalized simply as $\tilde M_{x, y} = M_{x, y} / L_x L_y$. Experimentally, the magnetotransport measurements are often expressed in terms of resistance. Here we first introduce the elementary AMR-induced resistance for a single vertex 
\begin{eqnarray}
	R_{0} = \frac{v}{i} = \frac{\alpha \Delta\rho}{t},
\end{eqnarray} 
where $v = \alpha \Delta \rho \, i / t$ is the parameter characterizing the voltage source introduced in Sec.~\ref{sec:v_source}; see also Fig.~\ref{fig:v_source}, $\alpha$ is a numerical constant depending on details of the geometry, and $t$ is the thickness of the film. For permalloy Ni$_{0.9}$Fe$_{0.1}$, the resistivity anisotropy is $\Delta \rho \approx 0.2$~$\mu\Omega\,$cm~\cite{perna14}. This corresponds to an elementary resistance $R_0 \approx 0.1\, \Omega$ for a $t = 20$ nm film. Normalized Hall resistance is then defined as $\tilde R_{\perp} = R_{\perp} / R_0$.  Using the relation $R_\perp = V_\perp / I$ and the total current $I = L_y \, i$, we have $\tilde R_\perp = V_\perp / (L_y \, i \, R_0) = \tilde V_\perp / v$, where the normalized Hall voltage is $\tilde V_\perp = V_\perp / L_y$. A similar definition is used for the AMR-induced correction to the longitudinal resistance, i.e. $\tilde R_\parallel = \Delta V_\parallel / I R_0 $, where $\Delta V_\parallel$ is the change of the longitudinal voltage. With our normalization convention, the maximum $\tilde V_{\perp, {\rm max}} = 2v$ (see Fig.~\ref{fig:m-rev-y}), giving rise to a maximum Hall resistance $R_{\perp, {\rm max}} = 2 R_0 \propto \Delta \rho / t$. Importantly, its independence on the length or width of the wires again indicates that the dominant contribution to $R_\perp$ is from the vertex regions. 

\subsection{Reversal of $x$-polarized state}

\begin{figure}[t]
\includegraphics[width=1.0\columnwidth]{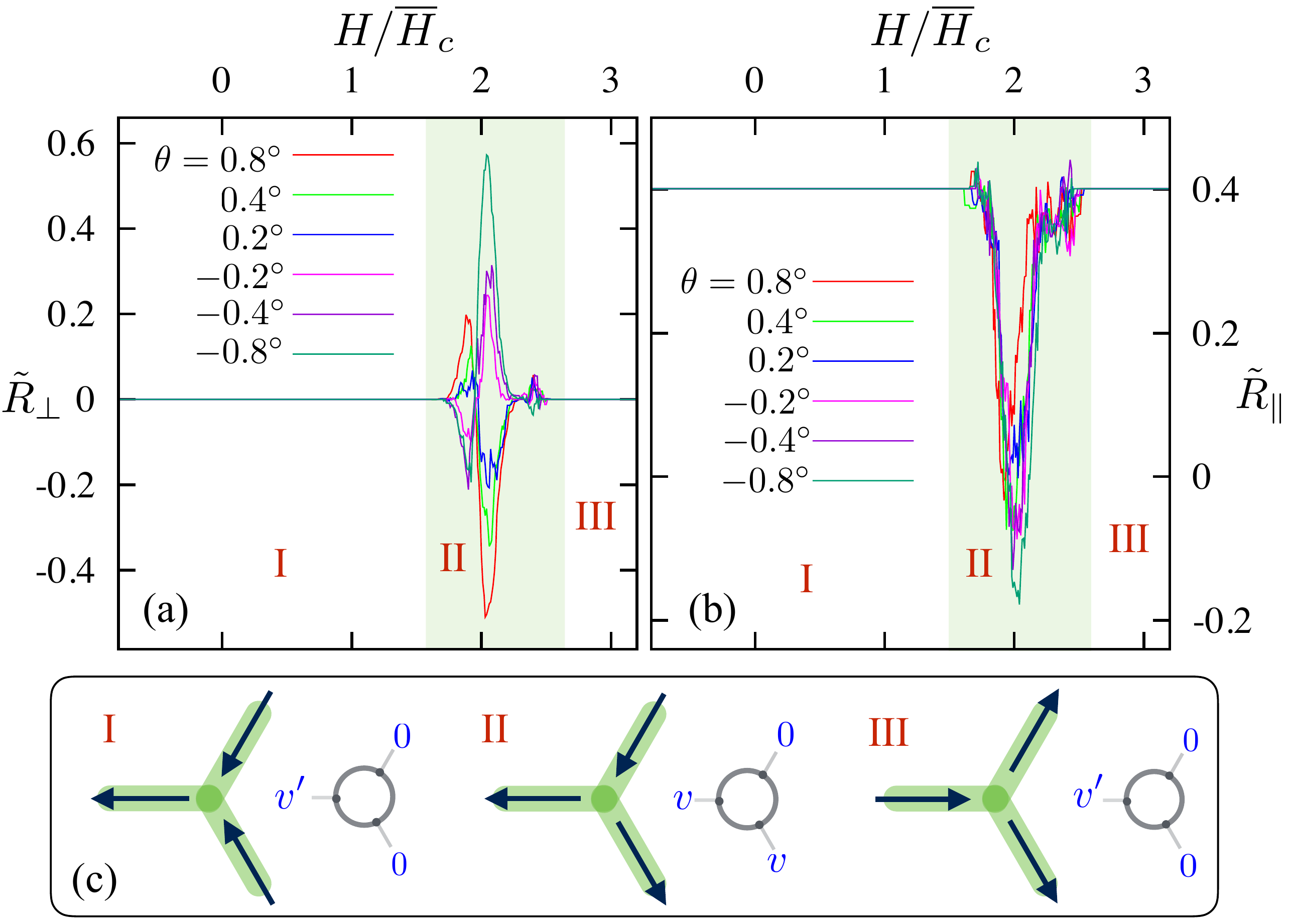}
\caption{(a) Transverse and (b) longitudinal resistances as functions of magnetic field during magnetization reversal for field direction close to the $x$ axis. The spin ice is initially in a polarized state with all spins pointing to the left. The reversal process can be divided into three regimes: the initial (I), the intermediate (II), and the final (III) regimes. Reversal of spins in the form of avalanches takes place in regime II. The representative vertices and their corresponding voltage source representation in the three regimes are shown in panel~(c).  Here the resistances are normalized with respect to the elementary AMR-induced $R_0$ in a single vertex. For typical permalloy Ni$_{0.9}$Fe$_{0.1}$ of nanometer thickness, $R_0 \approx 0.1$~$\sim$~$1\, \Omega$. 
\label{fig:m-rev-x}}
\end{figure}

We first consider the simpler case of magnetization reversal in a polarized state with spins pointing along the $-x$ direction. Such a state can be prepared by applying a strong magnetic field along the $-x$ direction. The AMR-induced Hall and longitudinal resistances as functions of the field strength are shown in Fig.~\ref{fig:m-rev-x}(a) and (b), respectively. The initial state is non-degenerate with all spins point to the left; this is labeled as regime I in Fig.~\ref{fig:m-rev-x}. By slowly increasing the magnetic field  within a small angle in the opposite $+x$ direction, the horizontal spins reach their coercivity field when $H \approx \overline{H}_c$. where $\overline{H}_c$ is the average coercive field. However, flipping of horizontal spins convert the 2-in-1-out vertices into higher energy three-in vertices which are prohibited by the ice rules. Consequently, magnetization reversal is initiated only when $H \sim  \overline{H}_c/\cos(60^\circ) = 2 \overline{H}_c$, which is the effective coercive field of the non-horizontal spins. This corresponds to regime II in Fig.~\ref{fig:m-rev-x}. The corresponding vertices in this regime exhibits a non-zero transverse voltage as shown in Fig.~\ref{fig:m-rev-x}(c), giving rise to a spike of the system Hall voltage whose sign depends on the small deviation of the field angle from the $+x$ direction. Finally, as $H$ is further increased, all spins are aligned by the magnetic field (regime III); the resultant polarized state again has zero Hall voltage. It is worth noting that such interesting spikes in Hall signal has been observed in experiments with similar geometrical setup~\cite{le16b}.

\begin{figure}[t]
\includegraphics[width=0.75\columnwidth]{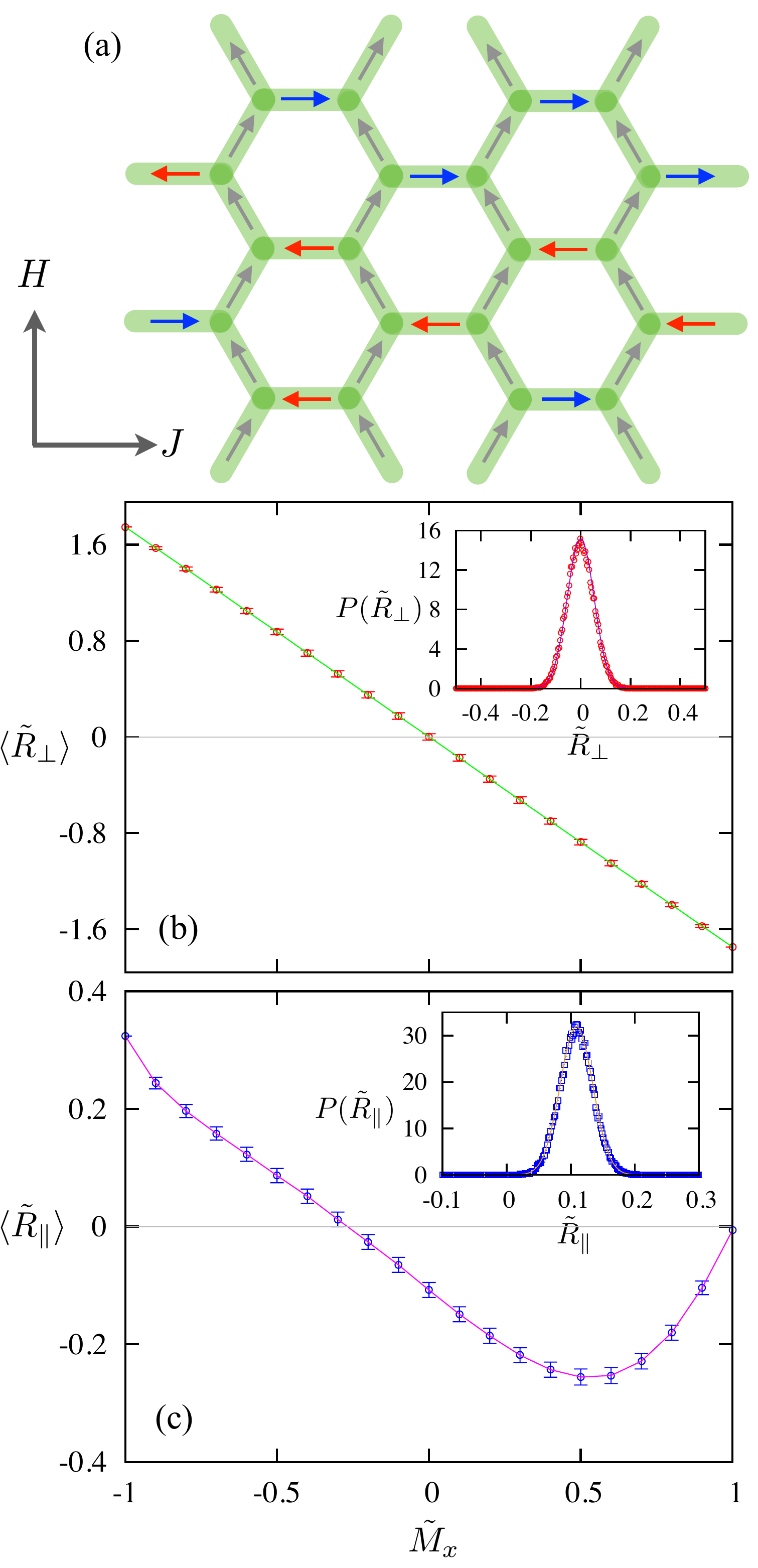}
\caption{(a) A partially ordered ice state with all non-horizontal spins point upward, while the horizontal spins remain disordered. The averaged AMR-induced (b) Hall and (c) longitudinal resistances as functions of the magnetization along $x$ direction. The curves are obtained by averaging over 5000 independent ice configurations satisfying $M_x = $ constant. The error bars correspond to the standard deviation of the finite distribution. The insets in (b) and (c) show the distribution of the transverse and longitudinal voltages, respectively, for state with $M_x = 0$.
\label{fig:y-polarized}}
\end{figure}

\subsection{Reversal of $y$-polarized state}
\label{sec:y-pol-state}

A simplified example of magnetization reversal induced by field applied along the $y$ axis is shown in Fig.~\ref{fig:m-rev-y}. The initial state is a long-range ordered state with a nonzero magnetization in both $x$ and $y$ directions. Experimentally, a strong magnetic field in the $y$ direction not only polarizes the non-horizontal spins, but also introduces significant magnetization deformation of the horizontal wires. As the field strength is reduced, the horizontal spins first relax to either $+x$ or $-x$ directions due to the magnetostatic shape anisotropy. However, this relaxation process is stochastic and depends on the experimental details such as disorders of the nanoarrays. Consequently, when the field magnitude is reduced to zero, the system ends up in a {\em partially} ordered state with polarized spins in sublattices 2 and 3, coexisting with disordered horizontal wires; see Fig.~\ref{fig:y-polarized}(a).

Here we first study the magnetotransport properties of such partially ordered ice states with non-horizontal spins aligned along the $y$-direction. For a given magnetization $M_x$ along the $x$-direction, there exists a huge degeneracy of microscopic ice configurations. Fig.~\ref{fig:y-polarized}(b) and (c) show the averaged Hall and longitudinal resistances as functions of $M_x$ for such partially ordered states. These curves were obtained after averaging over 1000 different configurations on an array with 2562 spins; the voltages of each configuration were computed using the circuit model. For example, the insets of Fig.~\ref{fig:y-polarized}(b) and (c) show the distribution of the computed resistance obtained from these 1000 different samples. The standard deviation of these distributions are indicated by the error bars of the main curves. Overall, the standard deviation is rather small, indicating a very narrow distribution of these voltages in such states.  Interestingly, our calculation finds a simple linear relation between the Hall voltage and $M_x$. On the other hand, a non-monotonic dependence of longitudinal voltage on $M_x$ was obtained.

\begin{figure}[t]
\includegraphics[width=1.0\columnwidth]{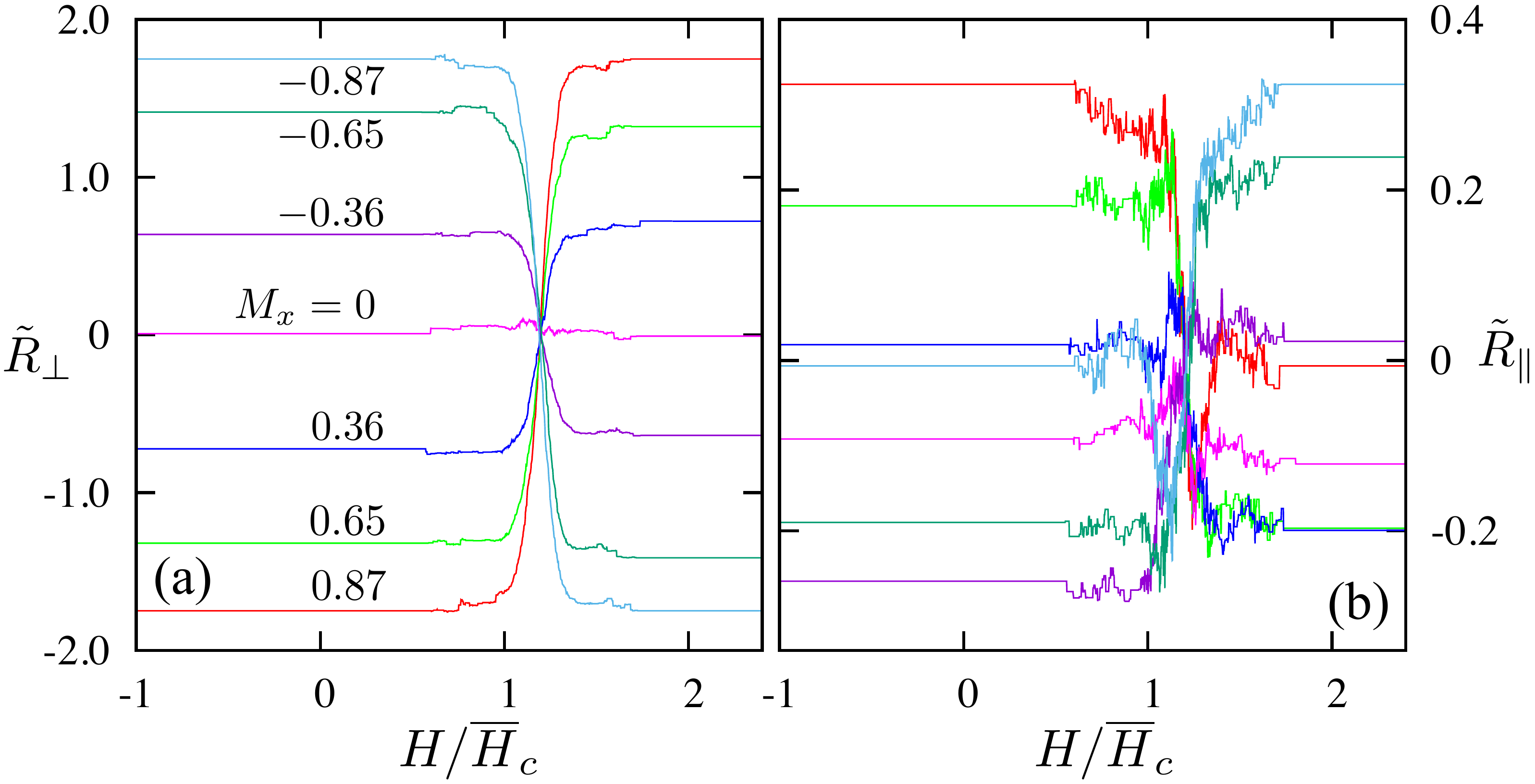}
\caption{(a) Transverse and (b) longitudinal resistances as functions of magnetic field during magnetization reversal for field direction close to the $y$ axis. The initial states have a maximum $M_y$ in the $-y$ direction, while the horizontal spins remain disordered. The different curves correspond to different values of $M_x$ of the particular spin configurations. Here the resistances are normalized with respect to the elementary AMR-induced $R_0$ in a single vertex. 
\label{fig:m-rev-y2}}
\end{figure}

The magnetotransport signals vs the reversal field strength are shown in Fig.~\ref{fig:m-rev-y2} for states with different $M_x$. Note that these curves were obtained for a single realization of the disordered horizontal spins. In the idealized situation, a magnetic field along the $y$-axis does not affect the horizontal spins (hence $M_x$ remains a constant), while the effective switching field for the non-horizontal wires is $H^{\rm eff}_c = \overline{H}_c / \cos(30^\circ) \approx 1.1547 \overline{H}_c$. Again, the magnetization reversal here proceeds through several spin avalanches at $H \sim H^{\rm eff}_c$. The Hall resistance flips sign after the reversal, consistent with the static property shown in Fig.~\ref{fig:y-polarized}(b). Moreover, the various $\tilde R_{\perp}$ curves show a clear crossing at the effective switching field. On the other hand, no universal behavior is seen for the longitudinal resistance, which appears to depend on the particular microscopic configurations.

\section{Mesoscopic constitutive relation}
\label{sec:m-reversal}

\subsection{Zero temperature result}

The linear dependence of the transverse resistance $R_\perp$ on the magnetization $M_x$ shown in Fig.~\ref{fig:y-polarized}(b) suggests a simple constitutive relation for the Hall voltage of the kagome spin ice. Indeed, by directly inspecting the vertices and the corresponding voltage sources shown in Fig.~\ref{fig:v_source}, one obtains the following relation between the Hall voltage and the total magnetization $\mathbf m_\alpha = (m_{\alpha, x}, m_{\alpha, y})$ of a vertex $\alpha$:
\begin{eqnarray}
	\label{eq:V_perp}
	v_{\alpha, \perp} =- \gamma \, \Delta \rho\, i \, m_{\alpha, x} \, m_{\alpha, y}.
\end{eqnarray}
Here $\Delta\rho = \rho_{\parallel} - \rho_{\perp}$ is the magnetoresistance anisotropy, $i$ is the current in a wire, and $\gamma$ is a numerical constant that depends on details of the geometry; see also Fig.~\ref{fig:v_source}(d). 

\begin{figure}[t]
\includegraphics[width=1.0\columnwidth]{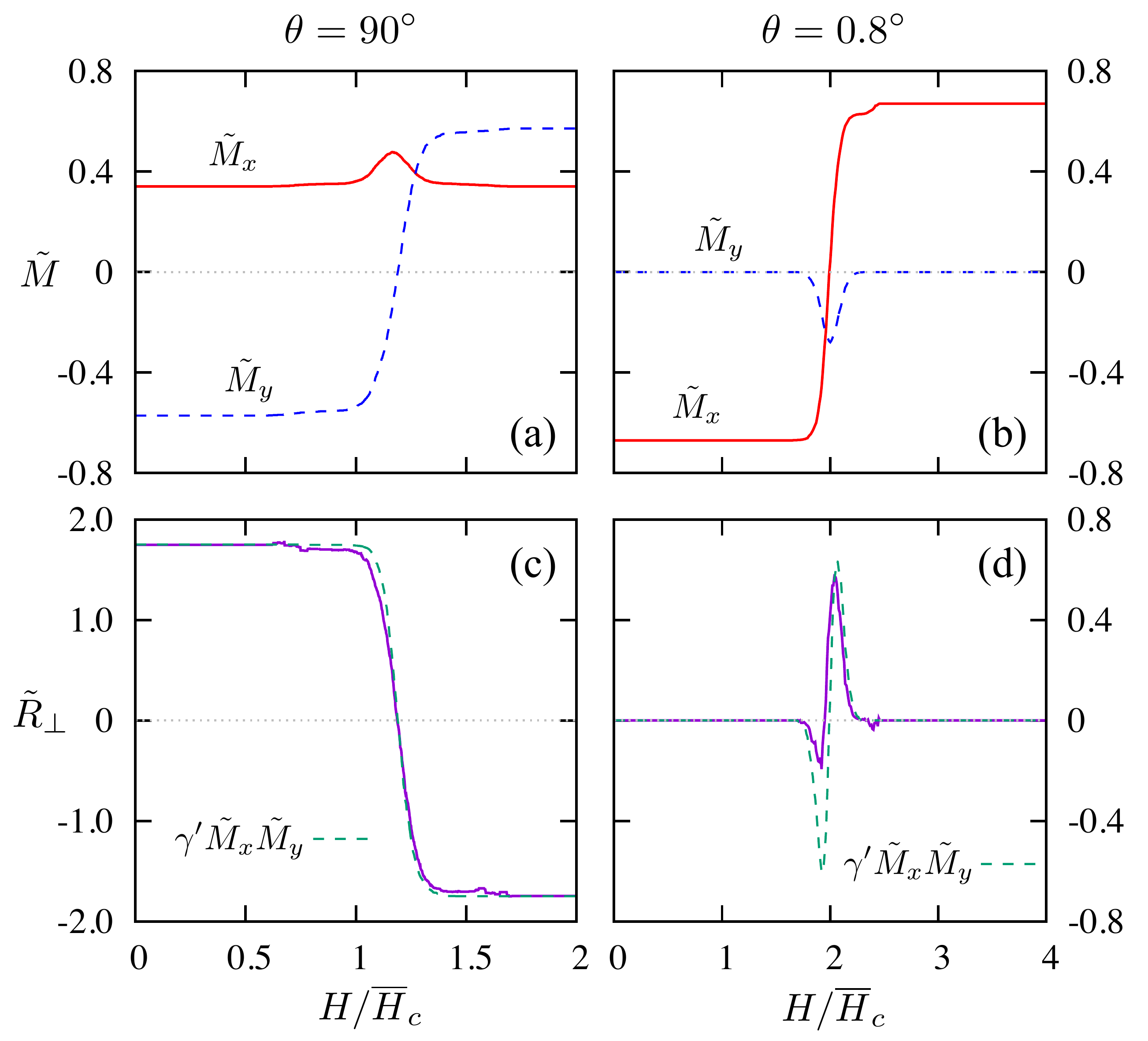}
\caption{Relaxation dynamics simulation of magnetization reversal in kagome spin ice. (a) and (b) show the magnetization curves with a magnetic field applied along direction $\theta = 90^\circ$ and $\theta=0.8^\circ$, respectively. The corresponding field dependence of the Hall voltage is shown in (c) and (d) for the two field angles, respecitvely. Here the same numerical constant $\gamma' =0.98$ is used all simulations.
\label{fig:mh}}
\end{figure}

Interestingly, this simple relation seems to hold also for the whole system. To demonstrate this, we first consider the evolution of Hall voltage during the magnetization reversal discussed in the previous section. Figs.~\ref{fig:mh}(a) and (b) show the $x$ and $y$ components of normalized magnetization during the reversal process for field angle $\theta = 90^\circ$ and $0.8^\circ$, respectively. A comparison between the magnetization and the Hall voltage curves clearly indicates that the sudden drop or spike of the Hall signal is intimately related to spin avalanches manifested as rapid increases of magnetization components.  The $\theta = 0.8^\circ$ case is a particularly telling example. The transverse voltage is nonzero (in the form of a spike) only when both $M_x$ and $M_y$ components are finite. Indeed, we find that the constitutive relation 
\begin{eqnarray}
	\label{eq:V_perp2}
	\tilde R_\perp \simeq -\Gamma R_0\, \tilde M_x \, \tilde M_y
\end{eqnarray}
provides a very good approximation for the whole lattice. Here $\Gamma$ is a numerical constant depending on details of the geometry, and $R_0$ is the elementary AMR-induced Hall resistance in a single vertex. For example, Hall resistances computed using this mesoscopic constitutive equation, shown as dashed lines in Figs.~\ref{fig:mh}(c) and (d), agree very well with those obtained from the circuit model. 

The Eq.~(\ref{eq:V_perp2}) should be viewed as an empirical relation which is valid only in the statistical sense. In the presence of a large magnetic field, the huge degeneracy associated with ice rules is lifted and the system is dominated by one particular type of vertex (depending on the field angle). Since $\tilde{\mathbf M} \approx \langle \mathbf m_\alpha \rangle $, and $\tilde V_\perp \approx \langle v_{\alpha, \perp} \rangle$ for such states, the mesoscopic relation follows directly from Eq.~(\ref{eq:V_perp}).  As demonstrated in Fig.~\ref{fig:y-polarized}(b), the above mesoscopic constitutive equation also describes the Hall voltage of the partially polarized states very well. Since in such states, $\tilde M_y = m_y$ is the same for all vertices, the transverse voltage $V_\perp = \sum'_\alpha v_{\alpha, \perp} = -\gamma \Delta \rho\, I m_y \sum'_\alpha m_{\alpha, x}$, where the prime indicates the summation is over a series of vertices from top to the bottom of the array. Eq.~(\ref{eq:V_perp2}) is then obtained noting that $\sum'_\alpha m_{\alpha, x} = 2 L_y \langle m_{\alpha, x} \rangle = 2 L_y \tilde M_x$. Of course, such states are still special case of the general spin configurations. From this simple example, the validity of Eq.~(\ref{eq:V_perp2}) relies on the property $\langle m_{\alpha, x} m_{\alpha, y} \rangle \approx \langle m_{\alpha, x} \rangle \langle m_{\alpha, y} \rangle$, which is not true in general, especially at finite temperatures. 


\subsection{Effects of finite temperature and Monte Carlo simulations}

 To examine the temperature effects, here we perform Monte Carlo simulations of the kagome spin ice at different temperatures. Specifically, we consider the following dumbbell model for kagome spin ice~\cite{castelnovo08,chern11}:
\begin{eqnarray}
	\label{eq:H-ice}
	\mathcal{H} = \sum_\alpha \frac{Q_\alpha^2}{2C} + \frac{D}{2}\sum_{\alpha \neq \beta} \frac{Q_\alpha \,Q_\beta}{|\mathbf r_\alpha - \mathbf r_\beta|} - \mathbf H \cdot \sum_\ell \mathbf m_\ell.
\end{eqnarray}
Here $Q_\alpha = \pm \sum_{\ell \in \alpha} \sigma_\ell$ is the net magnetic charge at vertex $\alpha$, the plus sign is used for one sublattice of vertex while the minus is for the other sublattice. The first term in the above expression represents a charging energy for each vertex, and is minimized by minimum charges at each vertex. Since each vertex is attached to three spins, the minimum charge $Q_{\rm min} = \pm 1$, corresponding to the pseudo-ice rules of kagome spin ice. In terms of Ising spins, the first term in Eq.~(\ref{eq:H-ice}) is equivalent to nearest-neighbor spin interaction $J_1 \sum_{\langle \ell m \rangle} \sigma_\ell \sigma_m$, with an interaction constant $J_1 = 1/C$. The second term in Eq.~(\ref{eq:H-ice}) denotes the Coulomb interaction between the vertex charges. Finally, the last term denotes the Zeeman coupling between nano-wires and the applied magnetic field. As discussed above, the magnetization of the nano-wire is $\mathbf m_\ell = \sigma_\ell m_0  \hat{\mathbf e}_{s_\ell}$.

The energy scales of the above Ising Hamiltonian are mainly determined microscopically by the magnetostatic dipolar interaction. The leading term $J_1 \sim \mu_0 \mu_m^2 / d^3$, where $\mu_0$ is the magnetic permeability, $d$ is the spacing between nearest-neighbor wires, and $\mu_m = M_s V$ is the magnetic moment of nanowire with $M_s$ being the saturation magnetization and $V = w l t$ the volume of the nanowire. Consequently, the energy scale depends crucially on the dimension of nanowires as well as their spacing. For permalloy such as Ni$_{80}$Fe$_{20}$, current fabrication technology is able to achieve an nanoarray with $J_1 \sim 10^2$ --$ 10^3$~K~\cite{farhan16,zhang13}. Importantly, this energy scale, which is of similar order to the Curie and blocking temperature of the nanowire, allows the array to be effectively thermalized. The Coulomb term $D$ comes from higher-order terms of the multipole expansion~\cite{moller09}, its precise value depends on geometrical details. In general, its magnitude is a few fractions of the leading $J_1$.

\begin{figure}[t]
\includegraphics[width=0.99\columnwidth]{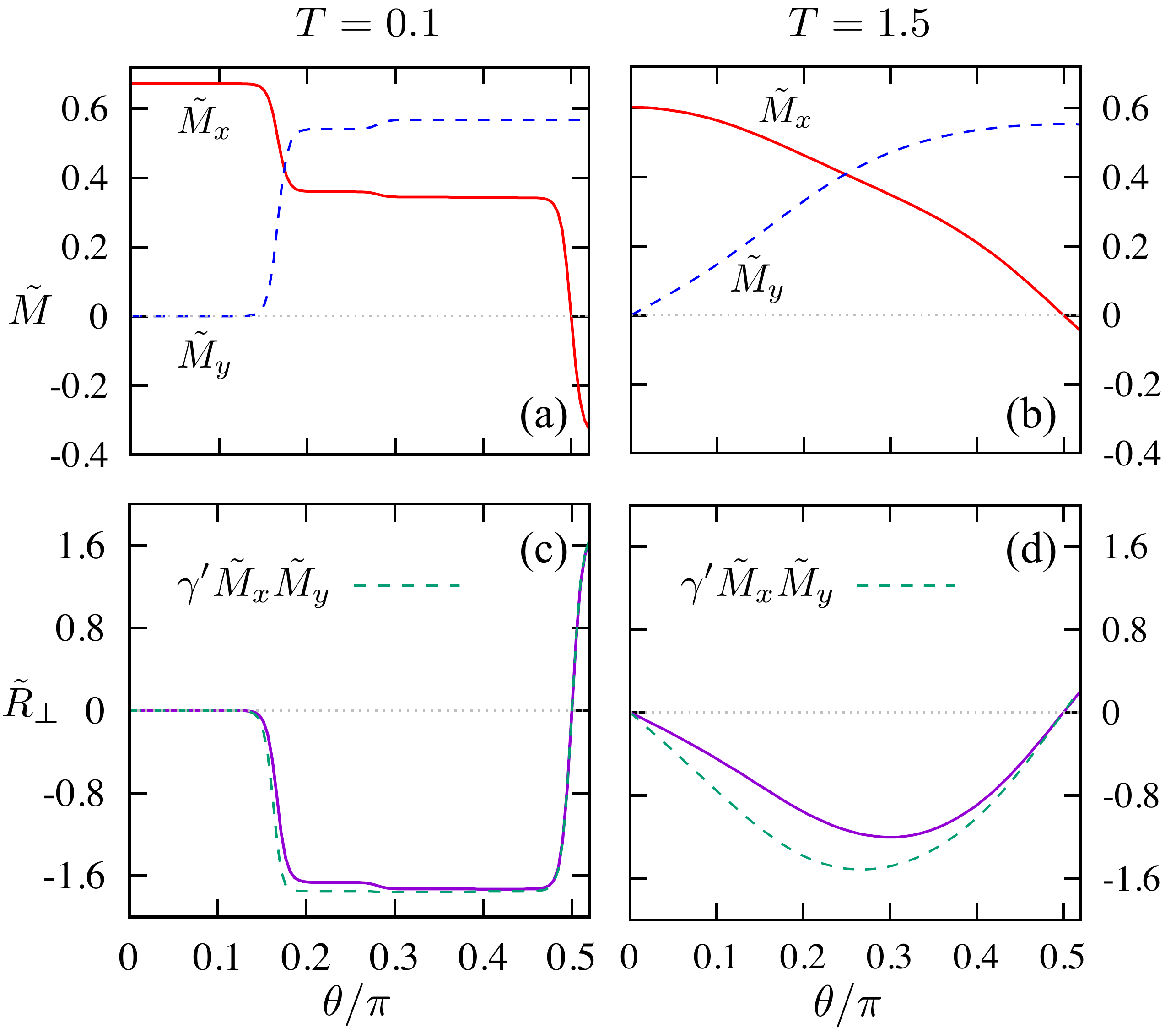}
\caption{Monte Carlo simulation kagome spin ice in the presence of a magnetic field. Equilibrium (normalized) magnetization $M_{x,y}$ as a function of field angle $\theta$ is shown in (a) and (b) for temperatures $T = 0.1 J_1$ and $T=1.5J_1$, respectively.  The corresponding field-angle dependence of the Hall resistance is shown in (c) and (d), where the (green) dashed lines are obtained by substituting $M_{x,y}$ into Eq.~(\ref{eq:V_perp2}).
\label{fig:m_theta}}
\end{figure}

We employ the Metropolis method in our Monte Carlo simulations of the dumbbell model Eq.~(\ref{eq:H-ice}). From snapshots generated during the simulations, the circuit model is used to compute the thermal averaged Hall voltage $V_\perp$. Fig.~\ref{fig:m_theta} shows the field-angle dependence of normalized magnetization  and transverse resistance $R_\perp$ of the network at two different temperatures $T = 0.1 J_1$ and $1.5 J_1$. Also shown for comparison is the Hall voltage computed using the formula Eq.~(\ref{eq:V_perp2}). A magnetic field $H = 3 J_1$ is applied in these simulations. As expected, the empirical relation Eq.~(\ref{eq:V_perp2}) provides a good approximation at low temperatures. In particular, the Hall voltage remains zero at small angles when no magnetization in the $y$-direction is induced by the field. Moreover, both $M_x$ and $R_\perp$ quickly drop to zero when the field is applied along the $y$-direction ($\theta \sim \pi/2$). On the other hand, significant deviation from the prediction of Eq.~(\ref{eq:V_perp2}) can be seen in the simulations at high temperatures. Nonetheless, the rule of thumb is that a nonzero Hall signal in general requires both $M_x$ and $M_y$ to be finite.  As shown in Fig.~\ref{fig:m_theta}(b) and (d), due to thermal fluctuations, coexistence of $M_x$ and $M_y$ persists in most of the field angles, giving rise to a finite Hall voltage except at  $\theta = 0$ and~$90^\circ$.

We conclude this section by noting that even though condition of a dominant vertex-type is not satisfied at small field or higher temperatures, the relation Eq.~(\ref{eq:V_perp2}) still provides a useful guideline for understanding the transverse voltage.

\begin{figure}
\includegraphics[width=1.0\columnwidth]{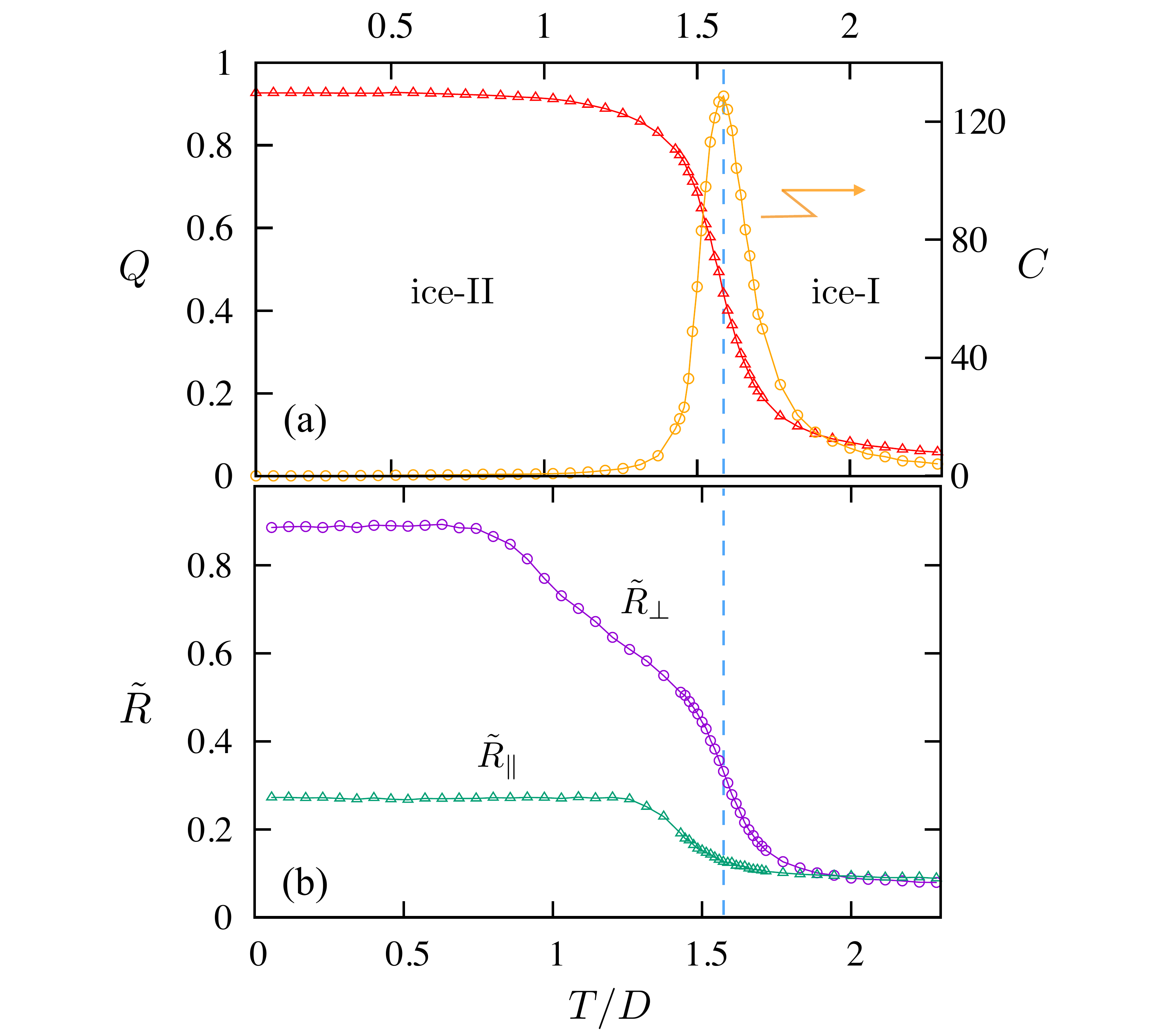}
\caption{Monte Carlo simulations of magnetic charge ordering in kagome spin ice. (a) N\'eel-type magnetic charge order parameter $Q \equiv \langle Q_\alpha \rangle_A - \langle Q_\beta \rangle_B$ and specific heat $C$ as a function of temperature. A small magnetic field is applied along $y$ direction to induce a small magnetization $M_y$. (b) the enhancement of Hall resistance by magnetic charge ordering. The charge-ordering transition separates the low-temperature correlated regime of kagome spin-ice into the ice-I and ice-II phases.
\label{fig:mc}}
\end{figure}

\section{Magnetic charge ordering}

As discussed in the Introduction, magnetic charges play an important role in the physics of spin ice. In particular, long-range ordering of uncompensated magnetic charges at vertices of kagome ice has been demonstrated both numerically~\cite{moller09,chern11} and experimentally~\cite{zhang13,canals16,dun16,paddison16}. A natural question to ask is what are the effects of magnetic charge ordering on the magnetotransport properties of kagome spin ice. To this end, we combine Monte Carlo simulations with circuit model to investigate the charge-ordering transition of the dumbbell model Eq.~(\ref{eq:H-ice}). We show that long-range charge order enhances the Hall voltage of the honeycomb array.  In order to have a finite Hall signal, a small magnetic field $H = 0.05 J_1$ is applied along the $y$-direction, giving rise to a small magnetization $M_y$. 

We first summarize the thermodynamic behaviors of kagome spin ice~\cite{moller09,chern11}. At temperatures $T \lesssim J_1$, the array enters a strongly correlated regime called the ice-I phase, in which short-range spin correlations are governed by the 2-in-1-out/1-in-2-out pseudo-ice rules. Configurations in this ice-I manifold is extensively degenerate with an entropy density $s_{\rm I} = 0.501$ per spin~\cite{wills02}. As temperature is further lowered, the system undergoes a continuous phase transition at $T \approx 1.55 D$ to the ice-II phase. The transition into the ice-II phase is marked by a pronounced peak in specific-heat $C$ and a rapid increase of the charge order parameter $Q$, see Fig.~\ref{fig:mc}(a). Here the charge order parameter $Q \equiv \langle Q_\alpha \rangle_A - \langle Q_\beta \rangle_B$ characterizes the staggered arrangement of magnetic charges on the two types of vertices in the honeycomb lattice~\cite{moller09,chern11}. 

In the ice-II phase, the sublattice symmetry is spontaneously broken such that vertices of one sublattice acquire a nonzero net magnetic charge (say $Q_A = +1$), while vertices of the other sublattice develops a net charge of opposite sign ($Q_B = -1$). This N\'eel type charge order results from the magnetic Coulomb interaction of Eq.~(\ref{eq:H-ice}). In terms of spins, the charge-order indicates that type-A (B) vertices are predominantly in one of the three 2-in-1-out (1-in-2-out) states. Crucially, as each vertex still has three possible configurations compatible with the charge constrains, spins remain disordered in the charge-ordered ice-II phase~\cite{moller09,chern11}. In fact, the ice-II phase contains extensively degenerate micro-states and is characterized by a nonzero entropy density $s_{\rm II} \approx 0.108 k_B$ per spin~\cite{moller09,chern11}.

\begin{figure}
\includegraphics[width=0.98\columnwidth]{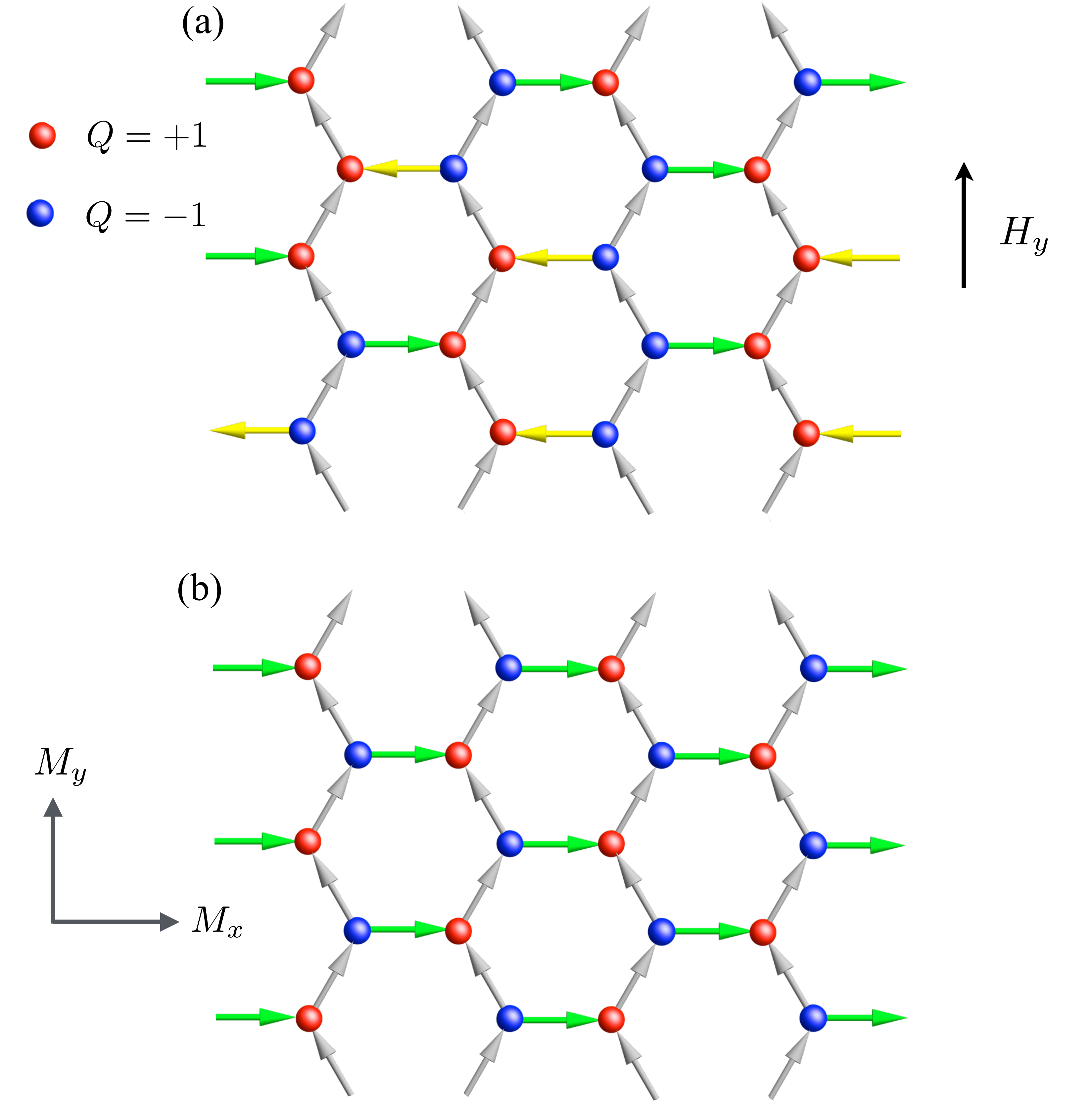}
\caption{Snapshots of ice-I and ice-II states in the presence of a strong magnetic field in the $y$ direction which polarizes the non-horizontal spins. (a) Disordered horizontal spins in the ice-I phase correspond to disorder of vertex magnetic charges. (b) the development of a staggered charge order in the ice-II phase induces a finite $M_x$, giving rise to a finite transverse voltage.
\label{fig:charge_order}}
\end{figure}

Applying a magnetic field $H_y$ partially removes the degeneracy of ice-I phase; two of the six 2-in-1-out/1-in-2-out vertices have a lower energy and remain degenerate. This residual degeneracy corresponds to the fact that horizontal spins are not affected by a magnetic field along $y$-direction. An special case is the manifold of partially ordered ice states discussed in Sec.~\ref{sec:y-pol-state}. In such a state shown in Fig.~\ref{fig:charge_order}(a), all non-horizontal spins are polarized by the field while the horizontal spins remain disordered. Consequently, there exists a macroscopic degeneracy $W_{\rm I'} = 2^{N_h}$ and a finite entropy density $s_{\rm I'} \approx 0.231 k_B$ per spin in this manifold, where $N_h = N/3$ is the number of horizontal wires. Although this special manifold of polarized states is stabilized only by a strong $H_y$ field, the important conclusion is that a residual entropy remains in the ice-I phase in the presence of a magnetic field in the $y$ direction. More importantly, the residual degeneracy indicates not only a vanishing $\langle M_x \rangle = 0$, but also disordered magnetic charnges $Q = 0$ in such ice-I phase; see Fig.~\ref{fig:charge_order}(a). 

As temperature is lowered, the array undergoes a charge-ordering transition, as in the case without the $H_y$ field, due to the Coulomb interaction between uncompensated magnetic charges. Crucially, the development of a long-range charge order below the phase transition induces a finite magnetization along the $x$-direction. This phenomenon is demonstrated in Fig.~\ref{fig:charge_order}(b) for the special case of $y$-polarized states.
Phenomenologically, the charge ordering transition of kagome spin ice is described by a Landau free energy expansion $\mathcal{F} = a (T - T_c) Q^2 + u\, Q^4 + \cdots$, where $a$ and $u$ are positive constants. It has been shown that the transition belongs to the 2D Ising universality class~\cite{chern11}. The enhancement of the Hall voltage can be understood by noting that the coupling term $Q M_x M_y$ is allowed in the Landau free energy expansion when time-reversal symmetry is explicitly broken by the external magnetic field $H_y$.  A nonzero $M_x \propto Q M_y$ thus results when the system enters the charge-ordered ice-II phase.  The presence of both $M_x$ and $M_y$ in turn produces a nonzero transverse voltage according to Eq.~(\ref{eq:V_perp2}). 


\section{Discussion and outlook}

To summarize, we have developed a theoretical model to understand the complex magnetotransport phenomena in artificial kagome spin ice. Microscopically, our theory is based on the anisotropic magnetoresistance (AMR) property of ferromagnetic permalloy islands that make up the nano-arrays. We derive a novel circuit model for computing both the transverse and longitudinal voltages  based on the leading order approximation of a perturbation calculation assuming a small resistance anisotropy.  In this picture, the kagome spin ice can be viewed as a resistor network driven by voltage sources that correspond to vertices of the honeycomb array. The differential voltage between terminals of the source is controlled by local spin ordering that is governed by the ice rules. The circuit model thus underscores the many-body origin of the Hall signal in kagome ice.

It is worth noting that the Hall voltage in our model originates microscopically from the AMR and can be viewed as the nano-array generalization of the so-called planar Hall effect (PHE). The fact that PHE is the dominating mechanism also manifests itself in the observed signature $\sin(2\phi)$ dependence of Hall signal at larger field~\cite{le16}. In principle, part of the Hall signal of spin ice could also come from the anomalous Hall effect~\cite{nagaosa10}, which depends on the out-of-plane magnetization component of the system. Since each vertex in the honeycomb array contains an odd number of edge defects, which can be viewed as fractional vortices~\cite{tchernyshyov05,pushp13}, the magnetization in the vertex region is in general non-coplanar. Conduction electrons propagating in such non-coplanar spin textures could acquire a nontrivial Berry phase that also contributes to the Hall conductivity~\cite{nagaosa10}. This is basically the microscopic mechanism of anomalous Hall effect. However, we expect this contribution to be much smaller compared with the PHE mainly becasue the perpendicular magnetization is much smaller than the in-plane components. 

We have combined the circuit model method with relaxation dynamics and Monte Carlo simulations of artificial spin ices. From such studies, we propose an empirical mesoscopic constitutive relation that relates the system Hall voltage to its total magnetization components. Our Monte Carlo simulations taking into account the Coulomb interaction of magnetic charges reveals that the Hall voltage is enhanced by the presence of a long-range charge order. This study exemplifies the intriguing effects of magnetic charges on the magnetotransport properties of artificial spin ice. 

It is worth noting that our analysis of the magnetotransport can be directly generalized to spin ices or magnetic nano-arrays of different geometries. The unusual magnetotransport behaviors observed in artificial spin ice make these magnetic metamaterials promising candidate for field sensors and other devices associated with AMR. The circuit model not only provides a useful picture for understanding the magneto-transport properties, but also serves as a guiding principle for designing artificial spin-ices as reconfigurable magneto-resistive devices. One of the novel features of artificial spin ice is the potential to manipulate these emergent magnetic charge degrees of freedom, for example, through local injection of domain walls into the nanowires~\cite{sentker14,burn14,burn17}. We expect our method provide a useful tool to further understand the intriguing interplay of magnetic charges and magnetotransport in artificial spin ice.




\bigskip

\begin{acknowledgements}
The author would like to thank Brian Le, Jungsik Park, Yuyang Lao, Joseph Sklenar, Cristiano Nisoli, and Peter Schiffer for collaborations on related works and numerous insightful discussions. The author also thanks R. Moessner and M. Udagawa for useful discussions on related topics of artificial spin ice.
\end{acknowledgements}

\end{document}